\documentclass[11pt,a4paper]{article}
\pdfoutput=1
\usepackage{jcappub}
\usepackage[utf8]{inputenc}
\usepackage[UKenglish]{babel}
\usepackage{hyperref}
\usepackage{amssymb}
\usepackage{aas_macros}
\usepackage{xspace}

\newcommand{\grmgrzobs}{>\hat{m}>\hat{z}}
\newcommand{\grmgrz}{>{m}>{z}}
\newcommand{\grmdv}{>mdV}

\newcommand{\fnl}{f_{\rm NL}}
\newcommand{\lcdm}{$\Lambda$CDM\xspace}
\newcommand{\fsky}{f_{\rm sky}}
\newcommand{\E}{\times10}
\newcommand{\mzero}{\left.m\right|_{0}}
\newcommand{\cardiff}{{School of Physics \& Astronomy, Cardiff University, 5
The Parade, Cardiff, CF24 3AA, United Kingdom}}
\newcommand{\helsinki}{{Department of Physics, University of Helsinki and
Helsinki Institute of Physics, P.O. Box 64, FIN00014 University of Helsinki,
Finland}}

\title{A consistent approach to falsifying \lcdm with rare galaxy clusters}
\author[a]{Ian Harrison}
\affiliation[a]{\cardiff}
\emailAdd{ian.harrison@astro.cf.ac.uk}
\author[b]{and Shaun Hotchkiss}
\affiliation[b]{\helsinki}
\emailAdd{shaun.hotchkiss@helsinki.fi}

\abstract{
We consider methods with which to answer the question ``is any observed galaxy 
cluster too unusual for \lcdm?'' After emphasising that many previous attempts 
to answer this question will overestimate 
the confidence level at which \lcdm can be ruled out, we outline a consistent
approach to these rare clusters, which allows the question to be answered. We define three statistical measures, each of
which are sensitive to changes in cluster populations arising from different
modifications to the cosmological model. We also use these properties to define the
``equivalent mass at redshift zero'' for a cluster -- the mass of an equally
unusual cluster today. This quantity is independent of the observational survey
in which the cluster was found, which makes it an ideal proxy for ranking the
relative unusualness of clusters detected by different surveys. These methods are then used on a
comprehensive sample of observed galaxy clusters and we confirm that all are
less than $2\sigma$ deviations from the \lcdm expectation. Whereas we have only
applied our method to galaxy clusters, it is applicable to any isolated,
collapsed, halo. As motivation for future surveys, we also calculate where in
the mass redshift plane the rarest halo is most likely to be found, giving
information as to which objects might be the most fruitful in the search for
new physics.
}


\arxivnumber{1210.4369}

\begin{document}
\maketitle

\section{Introduction}
The current concordance \lcdm cosmology makes (via the halo mass function) definite predictions for the abundance of gravitationally bound dark matter haloes and how that abundance evolves with redshift. These haloes are visible to us as galaxy clusters, both through their baryonic matter content in the form of galaxies and hot, diffuse gas and directly via strong and weak gravitational lensing events. Because of the halo mass function's sensitivity to cosmology, it has been useful in constraining parameters within \lcdm, principally the power spectrum normalisation $\sigma_8$ and total matter fraction $\Omega_m$, \cite{Mantz2010, PlanckCollaboration2013}. It has also been shown to be sensitive to many of the plausible modifications to concordance cosmology such as primordial non-Gaussianity \citep{Matarrese2000}, dark energy models \citep{Weller2002, Baldi2012} and modifications to gravity \citep{Schmidt2009,Ferraro2011,Lombriser2012}.

Ongoing surveys are revealing what are expected to be the most massive galaxy
clusters at ever increasing redshifts. These objects represent a
tantalising possibility: because the high-mass tail of the halo mass function
descends very steeply, the observation of even a single galaxy cluster which is
massive enough at a given redshift has the potential to contradict the
predictions of \lcdm with high significance.  Following early work \cite{Jimenez2009,Holz2012,Colombi2011}, a number of
statistical tests have been used to determine whether any of the clusters
observed by recent experiments are in tension with the standard model of
cosmology. These tests have fallen broadly into two groups: those which
consider explicitly how likely a given cluster is to appear in a \lcdm universe
(which we will refer to as `rareness' methods), and those which make use of
Extreme Value Statistics (EVS). EVS methods predict the probability distribution function for the mass of the most-massive cluster expected within
a given survey window and have generally found no tension with \lcdm
\citep{Waizmann2012, Harrison2012, Chongchitnan2012} when applied correctly to
an \emph{a priori} defined survey window. In contrast, the rareness methods have
chiefly considered the probability that a cluster could exist with both greater
mass and redshift than those observed, either calculating tension explicitly \cite{Jee2009, Cay'on2011, Jee2011, Jimenez2009, Hoyle2011, Enqvist2011} or by constructing `exclusion curves' in the mass-redshift plane \cite{Williamson2011, Brodwin2012, Menanteau2012, Menanteau2013}. However, it has been observed \cite{Hotchkiss2011} that these rareness analyses suffer from the (now continued) use of a biased statistical method, first introduced by \cite{Jimenez2009} and \cite{Holz2012}, which causes them to overestimate the amount of tension a particular observation may cause. Both \cite{Hotchkiss2011} and \cite{Hoyle2012} have analysed clusters using an un-biased rareness method, finding no tension with \lcdm in accordance with the EVS methods.

Nonetheless, the idea that observations of individual clusters are capable of
falsifying a given cosmological model remains a valid and intriguing
possibility, even if no tension currently exists. Here we provide a consistent method to calculate the rareness of observed
galaxy clusters, first quantifying how extreme each cluster is relative to
others, before using the methods advocated in \cite{Hotchkiss2011} to
determine the degree of tension they may cause with the predictions of \lcdm. Our work here goes beyond that of \cite{Hotchkiss2011}, first by the inclusion
of the effects of (cosmological) parameter uncertainty, secondly by the
inclusion of the effects of (mass) measurement uncertainty and thirdly by
examining a much more comprehensive list of clusters (including some discovered
after \cite{Hotchkiss2011} was published). The code used to calculate the results in this paper is available at \url{https://bitbucket.org/itrharrison/hh13-cluster-rareness/}, allowing our methods to be applied to new clusters observed in the future.

The paper is organised as follows. In section \ref{sec:probing:clusters} we first define the methodology we use to calculate the expected number of clusters in a given region of the mass-redshift plane, this includes our choice of halo mass function and cosmological parameters. Then, in section \ref{sec:previous} we review previous methods relating to galaxy clusters and reaffirm their problem of \emph{overestimating} tension with the cosmological model. In section \ref{sec:equally_rare} we define three statistical measures of rareness which are sensitive to different modifications to \lcdm. In section \ref{sec:calculating_rareness:location} we show where the rarest clusters are most likely to be observed, and in sections \ref{sec:calculating_rareness:parameteru} and \ref{sec:calculating_rareness:measurementu} we explain how observational uncertainties will be dealt with. Next, in section \ref{sec:equal_rareness:m0} we define the `equivalent mass at redshift zero', by which clusters from different surveys may be 
compared and ranked. Finally, we apply these methods to a large sample of observed clusters in section \ref{sec:current_rareness}. We both rank all the clusters according to their equivalent mass at redshift zero and, by making conservative assumptions about selection functions, we also set upper limits on the amount of tension these clusters provide with \lcdm predictions. In this section we also discuss the correspondance between our result and those gained using Extreme Value Statistics. Finally, in section \ref{sec:conclusions} we discuss our results and conclude.

\section{Cosmology with rare galaxy clusters}
\subsection{Galaxy cluster abundance and cosmology}
\label{sec:probing:clusters}
In the standard
cosmological model with Gaussian initial conditions and hierarchical structure
growth, high-mass galaxy clusters are expected to evolve from high peaks in the
initial cold dark matter (CDM) density fluctuations. The smallest scales
collapse first, before merging over time to form ever more massive CDM haloes,
into which baryons fall to form galaxy clusters. Consequently, high mass
clusters are expected to be very rare at early times, as reflected in the
exponential steepness of the halo mass function $n(m,z)$. The steepness of this
tail is also highly sensitive to the physical assumptions which go into the
initial conditions and dynamical evolution of the dark matter over-density
field, meaning the observation of even a single sufficiently extreme (in terms
of both its mass and redshift) cluster has the potential to provide strong
evidence against a particular cosmological model.

The number of galaxy clusters expected to occur in a survey window covering
fraction of the sky $\fsky$ and sensitive to clusters with masses between
$m_{min}$ and $m_{max}$ at redshifts between $z_{min}$ and $z_{max}$ is given
by the integrated product of the halo mass function and volume element within
this region:
\begin{eqnarray}
  \label{eqn:N_haloes}
  \langle N \rangle &=&
  \fsky \left[ \int^{z_{\rm max}}_{z_{min}} \int^{m_{max}}_{m_{min}} dz \, dM \,
        \frac{dV}{dz}\frac{dn(M, z)}{dM} \right]. 
\end{eqnarray}
In real surveys the
mass of a halo is not measured directly, but via proxies such as X-ray gas
temperature $T_{X}$, galaxy velocity dispersion $\sigma_v$ or the thermal
Sunyaev-Zel'dovich (tSZ) Compton-$y$. The realities of detecting these proxies
mean that real surveys are not typically mass limited (although tSZ surveys
approach this) and the use of absolute mass and redshift limits is a crude
approximation to the real selection function. However, in this paper we will
endeavour to be conservative with our approximate selection functions,
providing lower limits on cluster detection probabilities. The methodology
presented here can still be applied in the advantageous situation where the
full selection function is known, and our conclusions are expected to be
stable.

Throughout this work, the cosmology assumed is that described by the
WMAP7+BAO+H0 ML parameters \cite{Komatsu2011}. From these parameters
we calculate the linear matter power spectrum $P(k)$ using the numerical
Einstein-Boltzmann code CAMB\footnote{\url{http://camb.info}} and in turn the
variance $\sigma^2(m,z)$, smoothed with a top hat window function $W(k; m)$ and
evolved to a redshift of $z$ with the normalised linear growth function
$D_{+}(z)$
\begin{eqnarray}
  \label{eqn:sigma}
   \sigma^2(m,z) =
          D_{+}^2(z)\int_{0}^{\infty} \frac{dk}{2\pi} \, k^2 P(k) W^2(k; R).
\end{eqnarray}
The calculated $\sigma(m,z)$ is then used in the version of the Tinker halo mass function \cite{Tinker2008}:
\begin{eqnarray}
  \label{eqn:tinker_hmf}
  \frac{dn(m, z)}{dm} =  A \left[ \left(\frac{\sigma}{b}\right)^{-a} + 1\right]
                         e^{ -c/\sigma^2}
                         \frac{\bar{\rho}_{m,0}}{m}
                         \frac{d\mathrm{ln}(\sigma^{-1})}{dm}.
\end{eqnarray}
which includes parameters which evolve with redshift: $A = 0.186(1 + z)^{-0.14}$, $a = 1.47(1 + z)^{-0.06}$, $b = 2.57(1
+ z)^{-0.011}$, $c= 1.19$. This mass function has been well tested against
large, high-resolution N-body simulations and has become the most frequently
used in cosmological analyses.

\subsection{Comparison with previous analyses}
\label{sec:previous}
Many observable quantities are potentially available to classify galaxy clusters: halo mass, profile and concentration; redshift; population of
galaxies (and their type, colour etc); gas temperature and many others. Values of these observables can be combined to define a statistic and, when an observation of a cluster is made, the value of the statistic for that observation can be calculated. If we then wish to use this statistic to do inference on our cosmological model then we need to calculate the probablity distribution for this statistic. It is then straightforward to determine how unlikely/rare a particular cluster would be in \lcdm (and a given survey) by calculating the probability that \emph{any} cluster could be observed with a value that exceeds the measured value of the statistic. This probability to exceed (PTE) is a direct measure of the tension an observation provides with \lcdm. Here we summarise the previous work of \cite{Hotchkiss2011} considering correct and incorrect ways in which to calculate this tension.

Many previous analyses wished to quantify whether some observed clusters were too massive or formed too early for \lcdm. The statistic typically used in these analyses \cite{Jee2009, Cay'on2011, Jee2011, Jimenez2009, Hoyle2011, Enqvist2011} is the Poisson probability of observing at least one cluster (with observed mass $\hat{m}$ and redshift $\hat{z}$ denoted by hats) with both greater mass and redshift than the one which has been observed:
\begin{equation}
\hat{R}_{\grmgrzobs} = 1 - \exp\left(-\left< N_{\grmgrzobs} \right>\right),
\end{equation}
In these analyses the value of $\hat{R}_{\grmgrzobs}$ was taken, directly, as the degree of tension a cluster provides with \lcdm. However, as first pointed out by Fergus Simpson \footnote{\url{http://cosmocoffee.info/viewtopic.php?p=4932\&highlight=\#4932}} and later expounded in \cite{Hotchkiss2011}, using $\hat{R}_{\grmgrzobs}$ as a PTE will lead to incorrect conclusions because it ignores the fact that (observable) clusters at lower redshift and higher mass or higher redshift and lower mass would have values of this ${R}_{\grmgrzobs}$ statistic equal to or lower than what was observed. As explained in \cite{Hotchkiss2011} the true probability of an observation exceeding ${R}_{\grmgrzobs}$ is necessarily greater than the value of ${R}_{\grmgrzobs}$, meaning a low value of ${R}_{\grmgrzobs}$ is \emph{not} an uncommon property for the most extreme galaxy clusters expected in a \lcdm universe. The correct probability can be found be finding the line in the mass-redshift plane of clusters which have an 
equal $\left< N_{\grmgrzobs} \right>$ and calculating the probability of observing a galaxy cluster \emph{anywhere} above this line.

\begin{figure}
  \begin{centering}
    \includegraphics[width=0.5\textwidth]{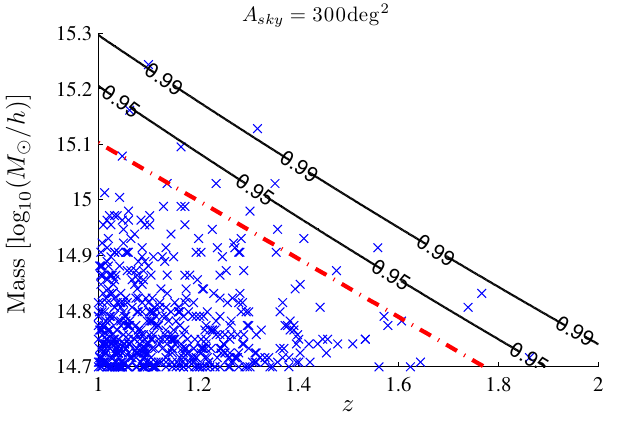}
      \includegraphics[width=0.5\textwidth]{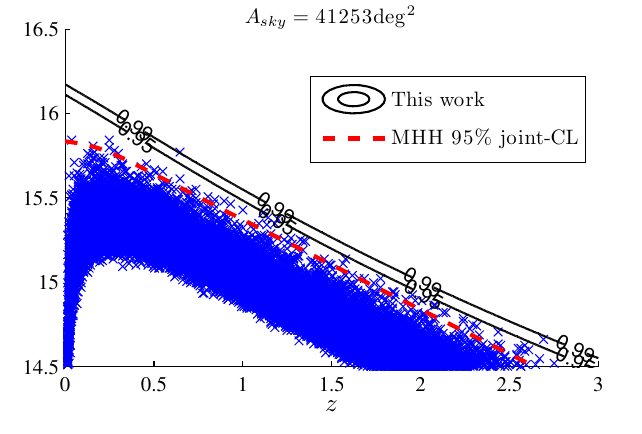}
  \end{centering}
  \caption{High-mass clusters from 100 Monte-Carlo realisations of the WMAP7 cosmology, plotted along with exclusion curves from Mortonson \emph{et al} (MHH) \cite{Mortonson2011} and this paper. As can be seen, significantly more than the expected $5$ clusters lie above the Mortonson \emph{et al} $95\%$ exclusion curve.}
  \label{fig:exclusion_curves}
\end{figure}

\begin{figure}
  \begin{centering}
    \includegraphics[width=0.5\textwidth]{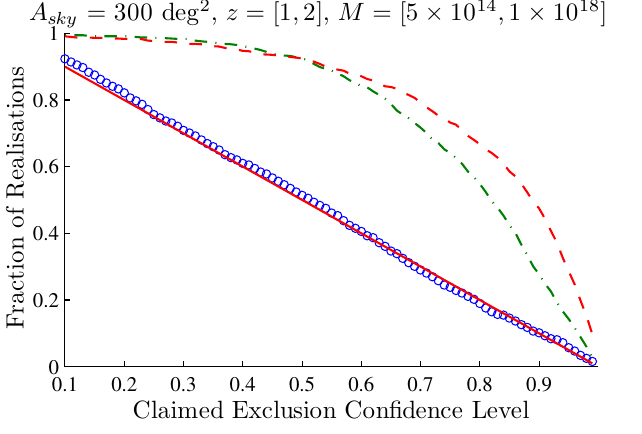}
    \includegraphics[width=0.5\textwidth]{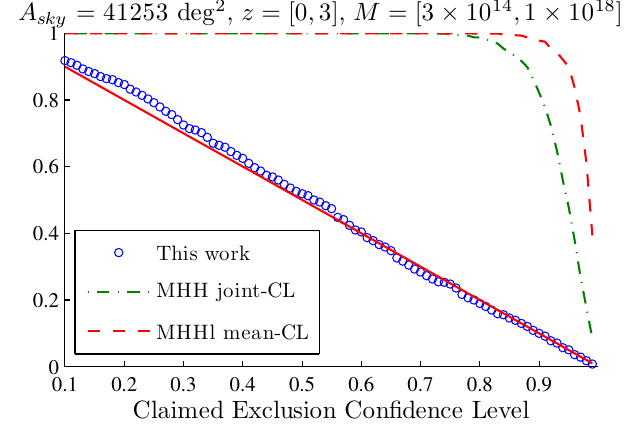}
  \end{centering}
  \caption{Fraction of WMAP7 cosmologies ruled out against confidence level, for both the Mortonson \emph{et al} (MHH) \cite{Mortonson2011} exclusion curves and those presented in this paper. Joint-CL refers to the choice $s=p$ in \cite{Mortonson2011} and Mean-CL $s=0.95, p=0.5$ (corresponding to sample
uncertainty only, with $\sigma_8$ and $\Omega_m$ at their maximum likelihood value). As can be seen, Mortonson \emph{et al} curves rule out the underlying cosmology to an erroneously high confidence level, with the correct behaviour represented by the solid straight line and followed closely by the results of this work.}
    \label{fig:sag_plots}
\end{figure}

This flaw in calibration also exists in the exclusion curves calculated by Mortonson \emph{et al} \cite{Mortonson2011} and hence in subsequent uses of these curves in the literature \cite{Williamson2011, Brodwin2012, Menanteau2012, Menanteau2013}. The defining property of an `exclusion curve' is that observation of a single cluster above the curve will rule out a \lcdm cosmology at the corresponding confidence level, meaning for an $100\alpha\%$ exclusion curve we should expect to observe a cluster above the line only $100(1-\alpha)\%$ of the time (i.e. because of a random fluctuation caused by sample variance). The curves from \cite{Mortonson2011} do not obey this property. Figure \ref{fig:exclusion_curves} shows 100 Monte-Carlo realisations of halo masses within a WMAP7 cosmology, along with a $95\%$ confidence level (CL) exclusion curve from \cite{Mortonson2011}. As can be seen, whilst there are $\sim 5$ clusters in the region $\grmgrz$ of each point on the line (as is to be expected from their construction), there are significantly 
more than the expected five clusters lying above the curve in total, a number which increases as more of the mass-redshift-$\fsky$ region is probed. Figure \ref{fig:sag_plots} further emphasises this; plotted is $\alpha$ against the fraction of Monte Carlo realisations of a \lcdm cosmology that contain a cluster that lies above a $100\alpha\%$ exclusion curve. It can clearly be seen that Mortonson \emph{et al} curves do not follow the behaviour required of correctly calibrated exclusion curves, represented by the solid straight line (i.e. that an $100\alpha\%$ CL-breaking cluster is found in $100(1-\alpha)\%$ of realisations). They instead show a significant hump, ruling out \lcdm at a high confidence level in a high fraction of realisations. Also displayed in figures \ref{fig:exclusion_curves} and \ref{fig:sag_plots} are the results gained using the analysis in this work, which do behave correctly as exclusion curves.

\section{Calculating the rareness of an observed cluster}
As discussed in section \ref{sec:previous}, we can correctly estimate the tension a galaxy cluster may be in with a given cosmological model, and define the related exclusion curves, by defining a rareness statistic, finding the contour of constant rareness and then calculating the probability of making an observation of a cluster anywhere above this line. As well as being correctly calibrated it is necessary to draw such curves in a physically motivated manner, as discussed in \cite{Hoyle2012}. Here, we identify three separate physically-motivated statistics which will later be used to calculate the PTE for observed clusters in a given cosmology and constuct correctly-calibrated exclusion curves.
\subsection{Three statistics to measure extremeness}
\label{sec:equally_rare}
\subsubsection{Expected number with greater mass and redshift $\grmgrz$}
\label{sec:equally_rare:grmgrz}
Even though it has been used incorrectly in previous works, the statistic defined by the expected number of clusters in a region with both greater mass and redshift:
\begin{equation}
\label{eqn:grmgrz}
  \langle N_{\grmgrz}\rangle =
               \left[ \int^{\infty}_{{z}} \int^{\infty}_{{m}} dz \, dM \,
               \frac{dV}{dz}\frac{dn(M, z)}{dM} \right].
\end{equation}
is intuitively physical and may be used in a correctly calibrated way, by finding the probability of observing a cluster anywhere above a line of constant $\left< N_{\grmgrzobs} \right>$. However, $\langle N_{\grmgrz}\rangle$ is sensitive to modifications in background expansion, growth and initial conditions, meaning well-motivated modifications to \lcdm are not separable.

\subsubsection{Expected number with greater initial peak height $>\nu$}
\label{sec:equally_rare:nu}
Galaxy clusters are expected to form at the location of high peaks in the distribution of primordial density perturbations, seeded by inflation. For a given fixed background expansion and growth law, changes in the CDM initial conditions, such as the widely-considered introduction of primordial non-Gaussianity (often parameterised by positive $\fnl$), would produce more rare clusters from higher peaks. We thus also consider the peak height from which a cluster is expected to have formed:
\begin{eqnarray}
  \nu(m,z) \propto \frac{1}{D_{+}(z)\sigma(m)},
\end{eqnarray}
as a physically-motivated rareness statistic.

\subsubsection{Expected number with greater mass, per unit volume $>mdV$}
\label{sec:equally_rare:grmdv}
Finally, we also use the statistic defined by the expected number of more massive
clusters per unit volume at a given redshift:\footnote{We thank Raul Angulo
(private correspondence) for motivating this definition.}
\begin{equation}
  \langle N_{>{m}dV} \rangle =
  \left[ \int^{\infty}_{{m}} dM \, \frac{dn(M, z)}{dM} \right].
\end{equation}
Using this definition has the advantage that it fairly weights all clusters at
high masses, even those which come from low-volume regions in the redshift
dimension.

\subsection{Expected masses and redshifts of the rarest clusters}
\label{sec:calculating_rareness:location}
\begin{figure}
  \begin{centering}
    \includegraphics[width=\textwidth]{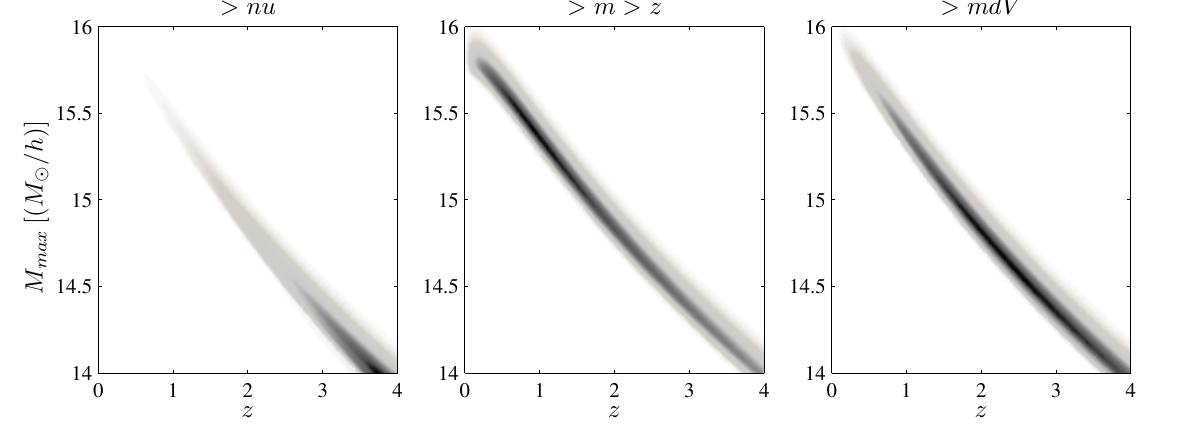}
    \caption{Heat map of the three
    statistics, showing where rarest clusters are most likely to be observed.}
    \label{fig:heats}
  \end{centering}
\end{figure}

We may also consider where in the mass-redshift plane we expect the rarest observed cluster to be found.
Answering this question can give information about where cluster surveys can be
most productively targeted, or indeed what kind of objects may be most
sensitive probes of the tail of the halo mass function. The plots in figure
\ref{fig:heats} show the probability distribution for the location in the mass-redshift plane of the rarest
observed cluster, for each statistic. The rarest cluster according to the $>\nu$ measure is always most likely to appear
at the highest specified redshift ($z=4$ for these plots), whilst the rarest
cluster according to the $>m>z$ and $>mdV$ measures are most likely to be
observed at $z\approx1$ and $z\approx2.5$ respectively.

An interesting inference can be made from the $>\nu$ plot with regards to
attempts to constrain primordial non-Gaussianity with rare objects. The
modification to the halo mass function caused by primordial non-Gaussianity
depends almost entirely on $\nu$. The tendency of surveys to be most likely to
find their rarest objects, according to the $\nu$ definitions, at the highest
possible redshift (and lower absolute masses) indicates that it is perhaps not
galaxy clusters but higher redshift events such as lensing arcs and quasars
which may prove the most sensitive probes of non-Gaussianity.

\subsection{Dealing with parameter uncertainty}
\label{sec:calculating_rareness:parameteru}
If we are seeking to test a cosmological model, it is necessary to take into account the uncertainties on the values
of the parameters within the model. As long as we do not introduce biases or make poor assumptions, we wish to be as sensitive to new physics as possible.
A statistically robust way to treat parameter uncertainty is to simply marginalise the probability to exceed $\hat{R}$ over available prior constraints on the cosmological parameters: \begin{eqnarray} \hat{R} = \int d\vec{\Lambda} \, \hat{R}(\vec{\Lambda})\Pi(\vec{\Lambda}), \end{eqnarray} where $\vec{\Lambda}$ is the full set of cosmological parameters and $\Pi(\vec{\Lambda})$ is the available prior probability distribution for those parameters. Of the standard model's cosmological parameters, it is the normalisation of the linear matter power spectrum $\sigma_8$ which has by far the most significant influence on cluster abundance. For the analysis below we use a Gaussian prior on $\sigma_8$ from \cite{Komatsu2011}, with a mean of $0.811$ and standard deviation of $0.03$.

\subsection{Dealing with measurement uncertainty}
\label{sec:calculating_rareness:measurementu}
A final consideration to be made when examining high-mass galaxy clusters is
the expected posterior distribution for the cluster mass $P(m|\hat{m})$, for which we follow the treatment of \cite{Andreon2009}. Here,
$\hat{m}$ is to be understood as the full set of observable parameters relating
to the measurement of a cluster's mass. In Bayesian reasoning, the posterior probability distribution function for the cluster mass $m$ in terms of an observable $\hat{m}$ is proportional to the product of the likelihood of the observation $L(\hat{m})$ and a prior probability distribution for mass $\Pi(m)$. Here, $L(\hat{m})$ is taken to be the observed cluster mass and error region, with either a normal or log-normal form. Because the prior distribution on cluster mass (the halo mass function) varies significantly over the width of this likelihood, its effect must be taken into account. This effect constitutes the classical Eddington bias
for number counts: because there are significantly more clusters in lower mass
bins which may upscatter into higher bins than there are high mass clusters to
scatter downwards, we must adjust our number counts accordingly. This allows us to calculate the posterior distribution for the true mass of a galaxy cluster:
\begin{eqnarray}
  P(m|\hat{m}) dm &\propto& \frac{dN(m,z)}{dmdz}L(\hat{m})dm.
\end{eqnarray}
The PTE values given here are then
calculated by marginalising over their values for the support of this
distribution. This method will give the correct posterior mass uncertainty for
a \lcdm prior if and only if the original quoted observable uncertainties are
the statistically correct posterior mass uncertainties obtained assuming a
uniform/no prior on cluster mass.

\section{Ranking clusters with equivalent mass at redshift zero}
Once a statistic has been defined, we can \citep[as suggested in][]{Hotchkiss2011} gain an intuitive understanding of how extreme an observed cluster is by calculating how massive a cluster at redshift zero would need to be in order to have the same value of this statistic. We will denote this by $\mzero$, the `equivalent mass at redshift zero'. Unlike the probability that a cluster could be detected in a given survey, $\mzero$ is an intrinsic property of each cluster and does not depend in any way on the depth, region or any other property of the survey it was selected from. This allows for a comparison (or even a ranking) of the extremeness of objects detected in different surveys and at different redshifts.

\begin{figure}
  \begin{centering}
    \includegraphics[width=\textwidth]{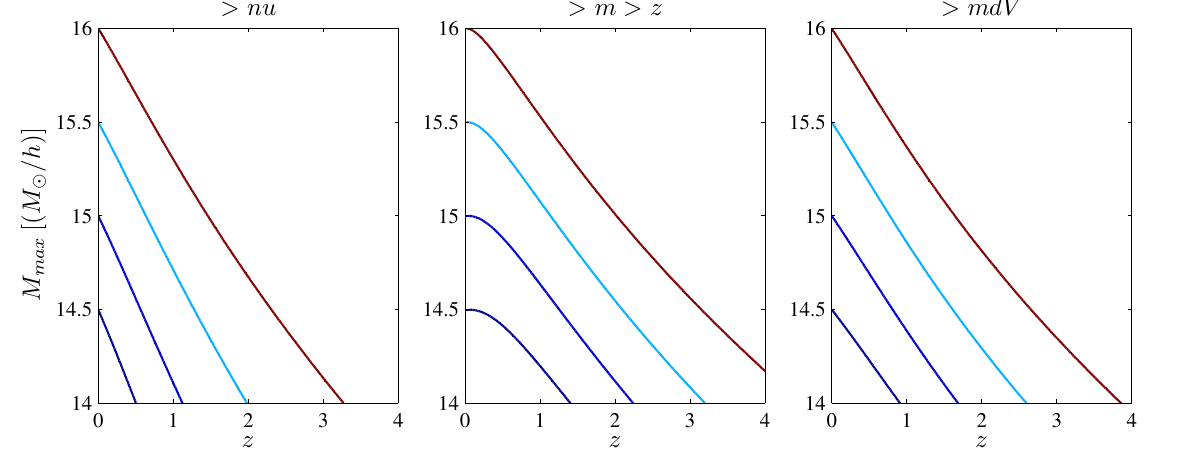}
    \caption{Contours of equal rareness, defined according to the three properties described in the text. For an observed cluster lying on each of these contours, the cluster's $\mzero$ is where the contour intersects the $z=0$ axis.}
  \label{fig:equal_contours}
  \end{centering}
\end{figure}

Figure \ref{fig:equal_contours} shows contours in the mass-redshift plane on which clusters will have equal values of the three statistics defined in section \ref{sec:equally_rare}. Where these contours intersect with the mass axis is $\mzero$. As can be
seen, the different definitions do not map points onto $\mzero$ in the same
way. For instance, the $\nu$ definition will assign the largest $\mzero$ to the
deepest fluctuation in the initial density field, irrespective of how the volume
expansion proceeds between that epoch and $z=0$, meaning they
appear as steeper contours on the mass-redshift plane. In contrast, the
tendency of the $\langle N_{\grmgrzobs} \rangle$ measure to downweight very low
redshift clusters because of the larger volume element at $z \lesssim 0.3$ can be seen in the flattening of the contours at these low redshifts.

\label{sec:equal_rareness:m0}

\section{Rareness and ranking of currently observed clusters}
\label{sec:current_rareness}
In this section we consider a large number of cluster observations and apply our methodology; we first calculate $\mzero$ for each to find which are the most extreme objects before finding the tension each observations represents with the standard cosmological model.

In order to calculate this tension we are required to find (as described in section \ref{sec:previous}) the probability for a particular observational survey to make any observation at least as rare as the detected galaxy clusters: the PTE. This requires knowledge of the survey selection function --- as a survey covers more of the sky and more of the mass-redshift plane it surveys more objects, increasing the number of objects which may be found with a given rareness. Here, we choose to conservatively set lower limits on the PTE (which correspond to upper limits on tension with \lcdm) by choosing suitable approximate selection functions. We do this by choosing the
minimal survey window in mass-redshift space in which the cluster may have been
found, as defined in table \ref{tab:surveys}. We do this by considering only the complete (i.e. where the probability
of detection $\rightarrow 1$) region of the survey, choosing high values
of $m_{min}$ and low values of $z_{max}$ for each survey and only considering
$\fsky$ for that particular survey. We also assume that the probability a cluster could exist in a region of the mass-redshift plane to be Poisson distributed (a good approximation for very high-mass galaxy clusters). A more sophisticated analysis would be possible on a per-survey basis, taking into account the full selection functions, such as the one which has been carried out by \cite{Stalder2013}, who perform a correctly-calibrated rareness analysis using simulations of their observational survey to compare with the observed cluster.

In addition to this, estimation of cluster masses is a procedure fraught with uncertainty. It has been found both observationally
\citep[see][and references therein]{Rozo2012} and in N-body simulations
\citep{Angulo2012} that masses (and ordering of most-massive clusters)
estimated using different proxies are frequently inconsistent with each other.
Further uncertainty occurs when converting between mass definitions for
comparison with halo mass functions: both a halo profile (frequently NFW) and a
mass-concentration relation must be assumed, both of which must be calibrated
using N-body simulations. Such considerations are outside the scope of this
paper. Here we choose to search the literature for published
estimations of cluster masses and take them `at face value'. This choice
na\"{i}vely ignores differences between survey mass proxies and sensitivities,
which may in reality widen published error estimates, and all of our
conclusions are predicated on this na\"{i}vity. However, where robust estimates
on cluster mass and uncertainty are available our methods will remain robust.

\subsection{Cluster catalogue}
\label{sec:current_rareness:catalogue}
Table \ref{tab:catalogues} shows the list of papers used to construct our cluster catalogue. In total 2334 cluster mass estimations were included, where
measurements in multiple proxies were allowed. As mentioned above, the
mass uncertainties on each method were taken to be those given by each paper
and were assumed to be normally distributed where error regions were symmetric
and log-normally distributed when asymmetric. For the MCXC catalogue
\citep{Piffaretti2011}, where no error estimates are given, a log-normal error
distribution with $\sigma_{\ln m}=0.2$ was assumed, as is fairly typical for
X-ray observations of clusters.

All cluster masses are converted to $m_{200m}$ (the mass which is within the
cluster region $200$ times the average density of the Universe) assuming an NFW
halo profile, with a single concentration parameter $c$, which is calculated
using the concentration-mass relation of \cite{Duffy2008} and WMAP7+BAO+H0 ML
parameters from \cite{Komatsu2011}. Where multiple cluster mass estimations appeared in the top-ten of $\mzero$, the observation with the smallest error region was used.

\begin{table*}
  \caption{The approximate survey selection functions used to calculate PTE values for the cluster catalogue.}
  \label{tab:surveys}
  \begin{center}\begin{tabular}{lllll}
    \hline
    Survey & $A_{sky} [\mathrm{deg}^2$] & $m_{min} [M_{\odot}h^{-1}$] & $z_{min}$ & $z_{max}$ \\
    \hline
    ACT & 755 & $8\E^{14}$ & 0.3 & 6.0 \\
    SPT & 2500 & $8\E^{14}$ & 0.3 & 6.0 \\
    XMM & 80 & $4\E^{14}$ & 0.9 & 1.5 \\
    MACS & 22735 & $8\E^{14}$ & 0.3 & 0.7 \\
    WARPS & 72 & $8\E^{14}$ & 0.0 & 0.6 \\
    PLCK & 41253 & $1\E^{15}$ & 0.3 & 6.0 \\
    RDCS & 50 & $8\E^{14}$ & 0.05 & 0.8 \\
    LoCuSS & 32085 & $3\E^{14}$ & 0.15 & 0.3 \\
    \hline
  \end{tabular}\end{center}
\end{table*}
\begin{table*}
  \caption{Papers used to compile cluster catalogue. $N_{Cl}$ is the number of
  clusters contained within each paper and observable mass proxies are WL, Weak
  Lensing; $N_{200}$, cluster richness; $T_{X}$, X-ray gas temperature; $Y_{X}$,
  integrated X-ray flux; $Y_{SZ}$, integrated Compton-$y$; $\sigma_{v}$, velocity
  dispersion; $L_{X}$, X-ray luminosity and $\zeta$, SPT matched filter
  signal-to-noise.}
  \label{tab:catalogues}
  \begin{tabular}{llllll}
    \hline
    Reference & Selecting Survey (Proxy) & $N_{Cl}$ & Mass Proxy & Notes \\
    \hline
    \hline
    \cite{McInnes2009} & SPT (SZ) & 3 & WL & N/A \\
    \cite{High2010} & SPT (SZ) & 21 & $N_{200}$ & N/A \\
    \cite{vSuhada2010} & SPT (SZ) & 2 & $T_{X}$ & N/A \\
    \cite{Andersson2011} & SPT (SZ) & 15 & $T_{X}, Y_{X}$ & N/A \\
    \cite{Brodwin2010} & SPT (SZ) & 1 & $Y_{SZ}, Y_{X}, \sigma_{v}, N_{200}$ & SPT-CLJ0546-5345 \\
    \cite{Marriage2011} & ACT (SZ) & 23 & $L_{X}$ & N/A \\
    \cite{Foley2011} & SPT (SZ) & 1 & $\zeta, Y_{X}, T_{X}, \sigma_{v}$ & SPT-CLJ2106-5844 \\
    \cite{Stalder2013} & SPT (SZ) & 1 & $\zeta, T_{X}$ & SPT-CLJ0205-5829 \\
    \cite{ThePlanckCollaboration2011} & Planck (SZ) & 10 & $Y_{X}$ & N/A \\
    \cite{Reichardt2013} & SPT (SZ) & 224 & $\zeta$ & N/A \\
    \cite{High2012} & SPT (SZ) & 5 & WL & N/A \\
    \cite{Consortium2012} & AMI (SZ) & 2 & $Y_{SZ}$ & DM-GNFW masses used \\
    \cite{Hasselfield2013} & ACT (SZ) & 91 & $Y_{SZ}$ & $M_{500c}^{UPP}$ masses used \\
    \hline
    \cite{Jee2009} & XMMU (X-ray) & 1 & WL & XMMU J2235.3 -2557 \\
    \cite{Rosati2009} & XMMU (X-ray) & 1 & $T_{X}$ & XMMU J2235.3 -2557 \\
    \cite{Gobat2011} & XMMU (X-ray) & 1 & $L_{X}$ & Highest $z$ \\
    \cite{Piffaretti2011} & Multiple (X-ray) & 1743 & $L_{X}$ & MCXC \\
    \cite{Fassbender2011} & XDCP (X-ray) & 22 & $L_{X}$ & N/A \\
    \cite{Jee2011} & Multiple (X-ray) & 22 & WL & All at $z>1$ \\
    \cite{vSuhada2012} & XMM-BCS (X-ray) & 46 & $L_{X}$ & N/A \\
    \hline
    \cite{Okabe2010} & LoCUSS (WL) & 30 & WL & N/A \\
    \cite{Demarco2010} & SpARCS (Optical) & 3 & $\sigma_{v}$ & N/A \\
    \cite{Menanteau2010} & SCSO (Optical) & 105 & $N_{200}$ & N/A \\
    \cite{Brodwin2012} & IDCS (IR) & 1 & $Y_{SZ}, L_{X}$ & N/A \\
    \cite{Vulcani2012} & EDisCS (Optical) & 1 & $WL$ & N/A \\
    \hline
    \hline  
  \end{tabular}
\end{table*}

\subsection{Rarest and most-massive clusters}
\label{sec:current_rareness:clusters}
Tables
\ref{tab:raresnu}-\ref{tab:raresgrmdv} show the clusters with the ten highest
values of $\mzero$ calculated using each of the three statistics defined in
section \ref{sec:equally_rare} and the PTE for that cluster in the relevant survey. Even with our conservative treatment of selection functions, the lowest PTE value is found to be as large as
$0.07$, for the cluster CLJ1226+3332. Note, however, that we have examined eight independent surveys. The probability that the smallest PTE in \emph{all} eight surveys is greater than or equal to $0.07$ is given by $(1-0.07)^8=0.56$. Therefore, if we live in a \lcdm universe, there is at least a $44\%$ chance that the smallest PTE in our tables will be less than or equal to $0.07$. Even with our very conservative treatment of selection functions, designed to make clusters appear rarer than they actually are, none of the clusters or surveys we have considered indicate any tension with \lcdm.

\begin{figure}
  \begin{centering}
    \includegraphics[width=0.475\textwidth]{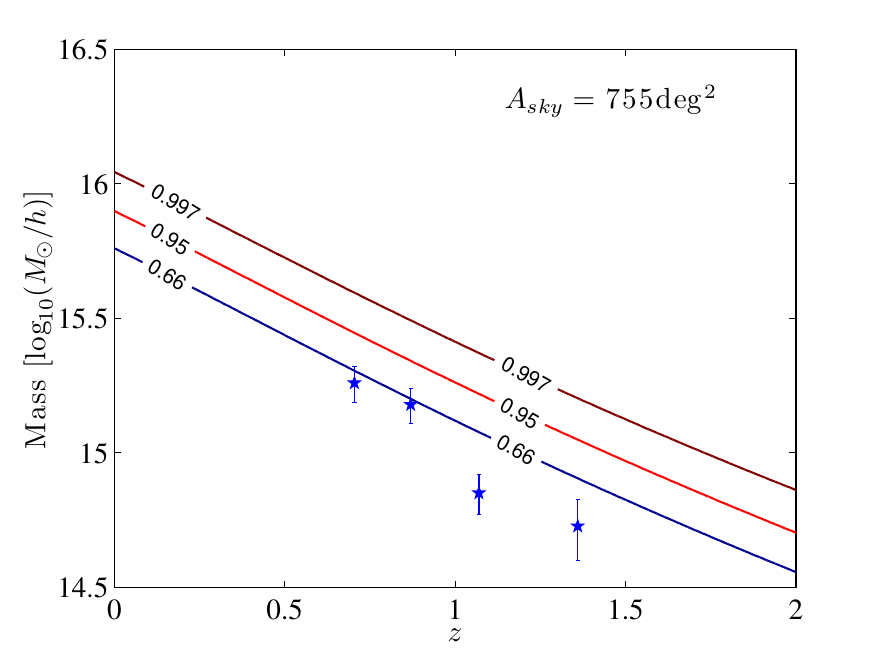}
    \includegraphics[width=0.475\textwidth]{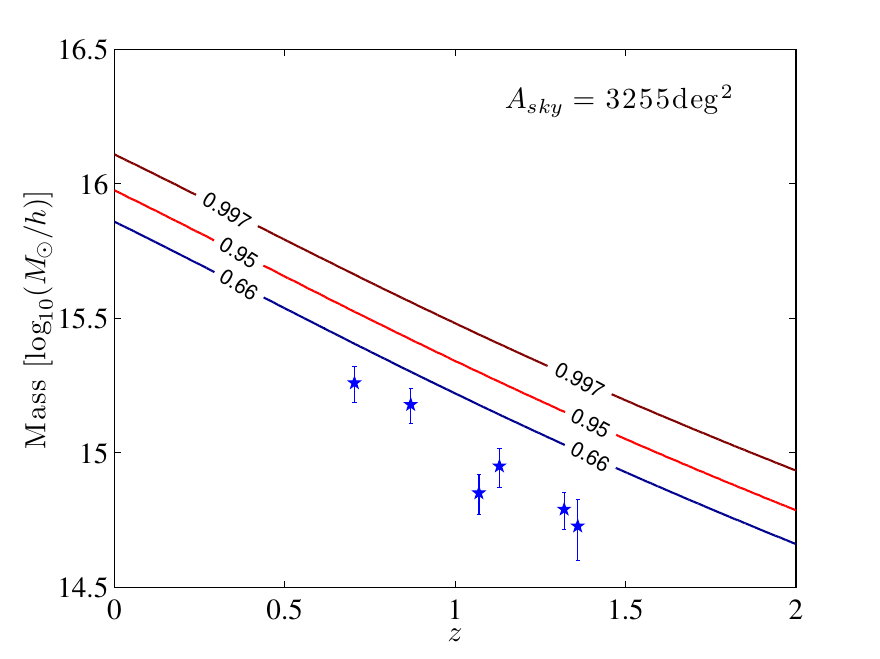}
    \caption{Currently observed clusters in tables \ref{tab:raresnu}-\ref{tab:raresgrmdv} and exclusion lines (using the $>mdV$ measure described in the text) corresponding to the ACT and ACT+SPT survey areas and sets of clusters.}
    \label{fig:clusterprobs}
  \end{centering}
\end{figure}

In order to demonstrate how these PTE relate to exclusion curves in figure \ref{fig:clusterprobs} we show the relevant plot for the ACT and SPT surveys. These correctly-calibrated exclusion curves were calculated using the ACT only and ACT+SPT survey regions and the clusters appearing in the top-ten tables are plotted. As can be seen, none of the clusters breaks the $66\%$ exclusion curve.

\begin{table*}
  \caption{The 10 clusters with highest $\mzero$ for the $\nu$ method described
  in the text. Masses marked with a * are uniform prior masses and were corrected using the procedure described in section \ref{sec:calculating_rareness:measurementu} before being used. All masses are in $10^{15}M_{\odot}h^{-1}$.}
  \label{tab:raresnu}
  \begin{tabular}{lllllll}
    \hline
    Cluster & $m_{200m}$ & z & Ref. (Proxy) & $\mzero^{\nu}$ & Survey & PTE \\
    \hline
    \hline
    ACT-CLJ0102-4915 & $1.51 \pm 0.22 $ & 0.87 & \cite{Menanteau2012} (Combined) & $6.53$ & ACT & 0.48\\
    ACT-CLJ2317-0204 & $1.82 \pm 0.28$ & 0.705 & \cite{Hasselfield2013} ($Y_{SZ}$) & $6.02$ & ACT & 0.60\\
    SPT-CLJ2106-5844 & $0.893\pm 0.148$ & 1.13 & \cite{Foley2011} (Combined) & $6.01$ & SPT & 0.85\\
    SPT-CLJ0205-5829 & $0.617\pm 0.096$ & 1.32 & \cite{Stalder2013} (Combined) & $5.70$ & SPT & 0.90\\
    ACT-CLJ0012-0046 & $0.534 \pm 0.137$ & 1.36 & \cite{Hasselfield2013} ($Y_{SZ}$) & $5.32$ & ACT & 0.73\\
    XMMUJ2235-2557 & $0.558_{-0.107}^{+0.129}$* & 1.39 & \cite{Jee2011} (WL) & $5.08$ & XMM & $0.40$\\
    MACSJ0417.5-1154 & $2.86_{-0.50}^{+0.62}$* & 0.44 & \cite{Piffaretti2011} $(L_{X})$ & $4.85$ & MACS & $0.97$\\
    CLJ1226+3332 & $1.12_{-0.16}^{+0.19}$* & 0.89 & \cite{Jee2011} (WL) & $4.55$ & WARPS & $0.07$\\
    ACT-CLJ0546-5345 & $0.709 \pm 0.121$ & 1.07 & \cite{Hasselfield2013} ($Y_{SZ}$) & 4.54 & ACT & 0.89\\
    PLCKG266.6-27.3 & $0.899\pm 0.067$* & 0.94 & \cite{ThePlanckCollaboration2011} $(Y_{X})$ & $4.49$ & Planck & $>0.99$\\
    \hline
    \hline
  \end{tabular}
\end{table*}

\begin{table*}
  \caption{The 10 clusters with highest $\mzero$ for the $\grmgrz$ method
  described in the text. Masses marked with a * are uniform prior masses and were corrected using the procedure described in section \ref{sec:calculating_rareness:measurementu} before being used. All masses are in $10^{15}M_{\odot}h^{-1}$.}
  \label{tab:raresgrmgrz}
  \begin{tabular}{lllllll}
    \hline
    Cluster & $m_{200m}$ & z & Ref. (Proxy) & $\mzero^{\grmgrz}$ & Survey & PTE \\
    \hline
    \hline
    ACT-CLJ0102-4915 & $1.51 \pm 0.22 $ & 0.87 & \cite{Menanteau2012} (Combined) & $3.43$ & ACT & 0.48\\
    ACT-CLJ2317-0204 & $1.82 \pm 0.28$ & 0.705 & \cite{Hasselfield2013} ($Y_{SZ}$) & $3.34$ & ACT & 0.50\\
    MACSJ0417.5-1154 & $2.86_{-0.50}^{+0.62}$* & 0.44 & \cite{Piffaretti2011} $(L_{X})$ & $3.03$ & MACS & $0.95$\\
    SPT-CLJ2106-5844 & $0.893\pm 0.148$ & 1.13 & \cite{Foley2011} (Combined) & $2.76$ & SPT & 0.92\\
    MACSJ2243.3-0935 & $2.23_{-0.39}^{+0.48}$* & 0.45 & \cite{Piffaretti2011} $(L_{X})$ & $2.45$ & MACS & $0.99$\\
    MACSJ2211.7-0349 & $2.36_{-0.41}^{+0.51}$* & 0.40 & \cite{Piffaretti2011} $(L_{X})$ & $2.42$ & MACS & $0.99$\\
    SPT-CLJ0205-5829 & $0.617\pm0.096$ & 1.32 & \cite{Stalder2013} (Combined) & $2.38$ & SPT & 0.97\\
    MACSJ0308.9+2645 & $2.15_{-0.38}^{+0.46}$* & 0.36 & \cite{Piffaretti2011} $(L_{X})$ & $2.23$ & MACS & $>0.99$\\
    ACT-CLJ0012-0046 & $0.534 \pm 0.137$ & 1.36 & \cite{Hasselfield2013} ($Y_{SZ}$) & $2.14$ & ACT & 0.89\\
    CLJ1226+3332 & $1.12_{-0.16}^{+0.19}$* & 0.89 & \cite{Jee2011} (WL) & $2.13$ & WARPS & $0.26$\\
    \hline
    \hline
  \end{tabular}
\end{table*}

\begin{table*}
  \caption{The 10 clusters with highest $\mzero$ for the $\grmdv$ method
  described in the text. Masses marked with a * are uniform prior masses and were corrected using the procedure described in section \ref{sec:calculating_rareness:measurementu} before being used. All masses are in $10^{15}M_{\odot}h^{-1}$.}
  \label{tab:raresgrmdv}
  \begin{tabular}{lllllll}
    \hline
    Cluster & $m_{200m}$ & z & Ref. (Proxy) & $\mzero^{>mdV}$ & Survey & PTE\\
    \hline
    \hline
    ACT-CLJ0102-4915 & $1.51 \pm 0.22 $ & 0.87 & \cite{Menanteau2012} (Combined) & $5.49$ & ACT & 0.46\\
    ACT-CLJ2317-0204 & $1.82 \pm 0.28$ & 0.705 & \cite{Hasselfield2013} ($Y_{SZ}$) & $5.20$ & ACT & 0.53\\
    SPT-CLJ2106-5844 & $0.893\pm 0.148$ & 1.13 & \cite{Foley2011} (Combined) & $4.70$ & SPT & 0.89\\
    MACSJ0417.5-1154 & $2.86_{-0.50}^{+0.61}$* & 0.44 & \cite{Piffaretti2011} $(L_{X})$ & $4.37$ & MACS & $0.96$\\
    SPT-CLJ0205-5829 & $0.617\pm 0.096$ & 1.32 & \cite{Stalder2013} (Combined) & $4.21$ & SPT & 0.95\\
    ACT-CLJ0012-0046 & $0.534 \pm 0.137$ & 1.36 & \cite{Hasselfield2013} ($Y_{SZ}$) & $3.84$ & ACT & 0.83\\
    MACSJ2243.3-0935 & $2.23_{-0.39}^{+0.48}$* & 0.45 & \cite{Piffaretti2011} $(L_{X})$ & $3.60$ & MACS & $0.99$\\
    CLJ1226+3332 & $1.12_{-0.16}^{+0.19}$* & 0.89 & \cite{Jee2011} (WL) & $3.60$ & WARPS & $0.13$\\
    XMMUJ2235-2557 & $0.558_{-0.107}^{+0.129}$* & 1.39 & \cite{Jee2011} (WL) & $3.58$ & XMM & $0.47$\\
    MACSJ2211.7-0349 & $2.36_{-0.41}^{+0.51}$* & 0.40 & \cite{Piffaretti2011} $(L_{X})$ & $3.48$ & MACS & $>0.99$\\
    \hline
    \hline
  \end{tabular}
\end{table*}

\subsection{Extreme Value Statistics of $\mzero$}
\label{sec:evs_m0}
\begin{figure}
  \begin{centering}
    \includegraphics[width=\textwidth]{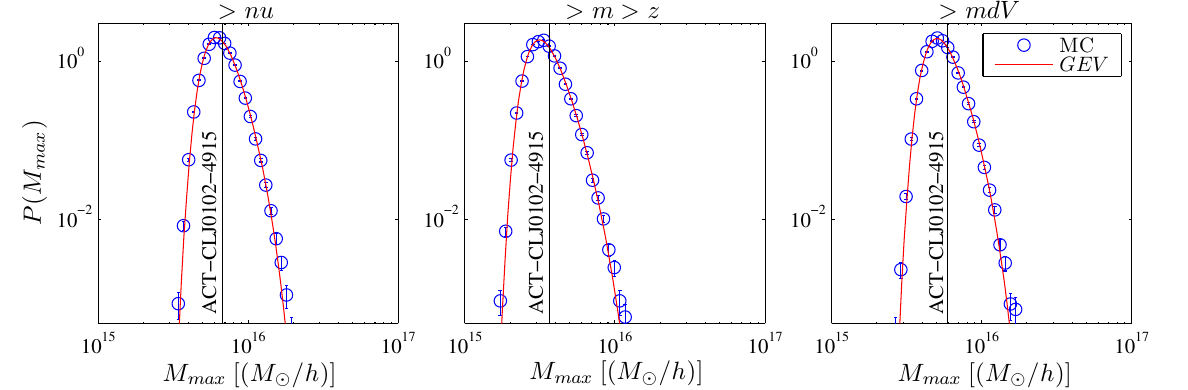}
    \caption{Extreme value distributions of $\mzero$ for the ACT survey definition in section \ref{sec:current_rareness:catalogue}. Blue points are $10^5$ Monte-Carlo realisations of the highest $\mzero$ in the survey, red lines are a GEV distribution fit to the points and the vertical line represents the mass of ACT-CLJ01012-4915, the highest $\mzero$ cluster in our tables.}
  \label{fig:evs_m0}
  \end{centering}
\end{figure}

Extreme Value Statistics (EVS) make predictions for the probability distribution function of sample extrema and have been used in the context of high-mass clusters by predicting the distribution for the most-massive cluster in a given survey region.
The PTEs calculated in section \ref{sec:current_rareness} represent the probability that at least one galaxy cluster exists in a survey region above a line of constant $\mzero$. This is $1-P_0$, where $P_0$ is the void probability that no clusters exist in the region above the line of constant rareness (constant $\mzero$). As emphasised by \cite{Davis2011}, this void probability is the same distribution as the EVS cumulative distribution function, the probability that the highest $\mzero$ in the survey region is less than or equal to the observed value. Forming the EVS from $\mzero$ contours in this way has the advantage of considering the entire survey region within a single distribution, unlike \cite{Harrison2012} which predicts the distribution for the most-massive cluster in narrow redshift bins with $\fsky=1$ or \cite{Waizmann2012} which considers smaller $\fsky$ but large redshift bins. However, the EVS distribution for $\mzero$ cannot be written down directly as it can be for the cluster mass $m$ only; here we obtain the distribution numerically by simulating $10^5$ highest $\mzero$s and fitting a Generalised Extreme Value (GEV, the limiting value for all extreme value distributions, see e.g. \cite{Gumbel1958}) distribution to the results. Figure \ref{fig:evs_m0} shows this procedure performed 
for the ACT survey selection function defined in table \ref{tab:surveys}, with the location of $\mzero$ for the most extreme object in the survey, ACT-CLJ0102-4915, also shown. As expected the probability for $\mzero$ to exceed the observed value on the EVS plot matches the PTE calculated in the rareness approach.

\section{Discussion and Conclusions}
\label{sec:conclusions}
In this paper we have considered an unbiased, consistent treatment of rare galaxy clusters. Because
previous considerations of cluster rareness have frequently fallen foul of
uncalibrated statistics that overestimate the amount of tension a given observation
is in with the \lcdm theory, we have been careful in defining the probabilities
we are calculating to avoid \emph{a posteriori effects}. We defined three statistics by considering three physically motivated properties of a cluster which may be sensitive to modifications in the underlying cosmology: the expected number of clusters at greater
mass and redshift; the peak height $\nu$ in the CDM over-density from which the
cluster grew; and the expected number of clusters with a greater mass, per unit volume. Using these statistics we calculated the probability that a defined survey would have observed a cluster as rare as an observed cluster or rarer, anywhere in the mass-redshift plane, i.e. the probability to exceed (PTE) the observed value of the statistic. This is a crucial difference to most earlier methods, wherein only clusters which had greater mass \emph{and} redshift were considered as more extreme than the one which had been observed.

We have also considered where in the mass-redshift plane the most extreme clusters
in a survey are most likely to reside. This provided us with an interesting
result: for the $\nu$ statistic, which is sensitive to primordial
non-Gaussianity, the most unusual cluster is always found at the highest
redshift available to the survey, meaning that, in principle, higher-redshift
objects (i.e. quasars, lensing arcs or gamma-ray bursts as opposed to galaxy
clusters) are potentially the more sensitive probes of non-Gaussianity in large
scale structure.

We also discussed a method to rank clusters between different surveys. This is $\mzero$, the equivalent mass at redshift zero. That is, the mass of the notional cluster at $z=0$ which has the same value of the chosen statistic, for an observed cluster. The value of $\mzero$ is an intrinsic property of
each cluster and does not depend at all on the survey in which it was found, meaning
it is an ideal proxy for categorising and ranking clusters according to their extremeness, even when they have been detected in different surveys and at different redshifts. In fact, this method is immediately generalisable to any isolated and collapsed halo for which a reliable mass measurement can be obtained, which would even allow us to compare and rank the relative ``extremeness'' of entirely different objects in a self-consistent way.

Finally, we have conducted a systematic review of cluster mass estimations in
the literature. Using conservative approximations to survey selection functions
and an `at face value' approach to published error estimates, we have
calculated the expected PTE for each cluster in its observational survey, finding that none are rarer
than the rarest cluster expected in some $7\%$ of surveys in \lcdm universes. As we have examined eight separate surveys, this value is entirely unremarkable.

To facilitate future estimates of galaxy cluster rareness we have made
a numerical code available at:
\url{https://bitbucket.org/itrharrison/hh13-cluster-rareness/}. This code will
calculate $m_{|0}$, PTE and a set of exclusion curves for any sets of
clusters that are subsequently observed.

\acknowledgments{
Ian Harrison receives an STFC studentship, thanks
Peter Coles and Chris Messenger for useful discussions and pays tribute to
Leonid Grishchuk for inspiring clear and rigorous science. Shaun Hotchkiss
acknowledges support from the Academy of Finland grant 131454. Both authors
thank Raul Angulo for motivating the $>mdV$ definition of extremeness used in
this paper.
}


\begin{thebibliography}{10}
\providecommand{\natexlab}[1]{#1}
\providecommand{\url}[1]{\texttt{#1}}
\providecommand{\urlprefix}{URL }
\providecommand{\eprint}[1]{\href{http://arXiv.org/abs/#1}{arXiv:#1}}

\bibitem{Mantz2010}
A.~{Mantz}, S.~W. {Allen}, D.~{Rapetti} et~al.
\newblock \emph{The observed growth of massive galaxy clusters - I. Statistical
  methods and cosmological constraints}.
\newblock \emph{\mnras} \textbf{406} (2010) 1759.
\newblock \eprint{0909.3098}.

\bibitem{PlanckCollaboration2013}
{Planck Collaboration}, P.~A.~R. {Ade}, N.~{Aghanim} et~al.
\newblock \emph{Planck 2013 results. XX. Cosmology from Sunyaev-Zeldovich
  cluster counts}.
\newblock \emph{ArXiv e-prints} \eprint{1303.5080}.

\bibitem{Matarrese2000}
S.~{Matarrese}, L.~{Verde} and R.~{Jimenez}.
\newblock \emph{The Abundance of High-Redshift Objects as a Probe of
  Non-Gaussian Initial Conditions}.
\newblock \emph{\apj} \textbf{541} (2000) 10.
\newblock \eprint{astro-ph/0001366}.

\bibitem{Weller2002}
J.~{Weller}, R.~A. {Battye} and R.~{Kneissl}.
\newblock \emph{Constraining Dark Energy with Sunyaev-Zel'dovich Cluster
  Surveys}.
\newblock \emph{Physical Review Letters} \textbf{88}~(23) (2002) 231301.
\newblock \eprint{astro-ph/0110353}.

\bibitem{Baldi2012}
M.~{Baldi}.
\newblock \emph{Early massive clusters and the bouncing coupled dark energy}.
\newblock \emph{\mnras} \textbf{420} (2012) 430.
\newblock \eprint{1107.5049}.

\bibitem{Schmidt2009}
F.~{Schmidt}, M.~{Lima}, H.~{Oyaizu} et~al.
\newblock \emph{Nonlinear evolution of f(R) cosmologies. III. Halo statistics}.
\newblock \emph{\prd} \textbf{79}~(8) (2009) 083518.
\newblock \eprint{0812.0545}.

\bibitem{Ferraro2011}
S.~{Ferraro}, F.~{Schmidt} and W.~{Hu}.
\newblock \emph{Cluster abundance in f(R) gravity models}.
\newblock \emph{\prd} \textbf{83}~(6) (2011) 063503.
\newblock \eprint{1011.0992}.

\bibitem{Lombriser2012}
L.~{Lombriser}, A.~{Slosar}, U.~{Seljak} et~al.
\newblock \emph{Constraints on f(R) gravity from probing the large-scale
  structure}.
\newblock \emph{\prd} \textbf{85}~(12) (2012) 124038.
\newblock \eprint{1003.3009}.

\bibitem{Jimenez2009}
R.~{Jimenez} and L.~{Verde}.
\newblock \emph{Implications for primordial non-Gaussianity ($f_{NL}$) from
  weak lensing masses of high-z galaxy clusters}.
\newblock \emph{\prd} \textbf{80}~(12) (2009) 127302.
\newblock \eprint{0909.0403}.

\bibitem{Holz2012}
D.~E. {Holz} and S.~{Perlmutter}.
\newblock \emph{The Most Massive Objects in the Universe}.
\newblock \emph{\apjl} \textbf{755} (2012) L36.
\newblock \eprint{1004.5349}.

\bibitem{Colombi2011}
S.~{Colombi}, O.~{Davis}, J.~{Devriendt} et~al.
\newblock \emph{Extreme value statistics of smooth Gaussian random fields}.
\newblock \emph{\mnras} \textbf{414} (2011) 2436.
\newblock \eprint{1102.5707}.

\bibitem{Waizmann2012}
J.-C. {Waizmann}, S.~{Ettori} and L.~{Moscardini}.
\newblock \emph{An application of extreme value statistics to the most massive
  galaxy clusters at low and high redshifts}.
\newblock \emph{\mnras} \textbf{420} (2012) 1754.
\newblock \eprint{1109.4820}.

\bibitem{Harrison2012}
I.~{Harrison} and P.~{Coles}.
\newblock \emph{Testing cosmology with extreme galaxy clusters}.
\newblock \emph{\mnras} \textbf{421} (2012) L19.
\newblock \eprint{1111.1184}.

\bibitem{Chongchitnan2012}
S.~{Chongchitnan} and J.~{Silk}.
\newblock \emph{Primordial non-Gaussianity and extreme-value statistics of
  galaxy clusters}.
\newblock \emph{\prd} \textbf{85}~(6) (2012) 063508.
\newblock \eprint{1107.5617}.

\bibitem{Jee2009}
M.~J. {Jee}, P.~{Rosati}, H.~C. {Ford} et~al.
\newblock \emph{Hubble Space Telescope Weak-lensing Study of the Galaxy Cluster
  XMMU J2235.3 - 2557 at $z \sim{} 1.4$: A Surprisingly Massive Galaxy Cluster
  When the Universe is One-third of its Current Age}.
\newblock \emph{\apj} \textbf{704} (2009) 672.
\newblock \eprint{0908.3897}.

\bibitem{Cay'on2011}
L.~{Cay{\'o}n}, C.~{Gordon} and J.~{Silk}.
\newblock \emph{Probability of the most massive cluster under non-Gaussian
  initial conditions}.
\newblock \emph{\mnras} \textbf{415} (2011) 849.
\newblock \eprint{1006.1950}.

\bibitem{Jee2011}
M.~J. {Jee}, K.~S. {Dawson}, H.~{Hoekstra} et~al.
\newblock \emph{Scaling Relations and Overabundance of Massive Clusters at $z
  \gtrsim 1$ from Weak-lensing Studies with the Hubble Space Telescope}.
\newblock \emph{\apj} \textbf{737} (2011) 59.
\newblock \eprint{1105.3186}.

\bibitem{Hoyle2011}
B.~{Hoyle}, R.~{Jimenez} and L.~{Verde}.
\newblock \emph{Implications of multiple high-redshift galaxy clusters}.
\newblock \emph{\prd} \textbf{83}~(10) (2011) 103502.
\newblock \eprint{1009.3884}.

\bibitem{Enqvist2011}
K.~{Enqvist}, S.~{Hotchkiss} and O.~{Taanila}.
\newblock \emph{Estimating $f_{NL}$ and $g_{NL}$ from massive high-redshift
  galaxy clusters}.
\newblock \emph{\jcap} \textbf{4} (2011) 17.
\newblock \eprint{1012.2732}.

\bibitem{Williamson2011}
R.~{Williamson}, B.~A. {Benson}, F.~W. {High} et~al.
\newblock \emph{A Sunyaev-Zel'dovich-selected Sample of the Most Massive Galaxy
  Clusters in the 2500 deg$^{2}$ South Pole Telescope Survey}.
\newblock \emph{\apj} \textbf{738} (2011) 139.
\newblock \eprint{1101.1290}.

\bibitem{Brodwin2012}
M.~{Brodwin}, A.~H. {Gonzalez}, S.~A. {Stanford} et~al.
\newblock \emph{IDCS J1426.5+3508: Sunyaev-Zel'dovich Measurement of a Massive
  Infrared-selected Cluster at z = 1.75}.
\newblock \emph{\apj} \textbf{753} (2012) 162.
\newblock \eprint{1205.3787}.

\bibitem{Menanteau2012}
F.~{Menanteau}, J.~P. {Hughes}, C.~{Sif{\'o}n} et~al.
\newblock \emph{The Atacama Cosmology Telescope: ACT-CL J0102-4915 ''El
  Gordo,'' a Massive Merging Cluster at Redshift 0.87}.
\newblock \emph{\apj} \textbf{748} (2012) 7.
\newblock \eprint{1109.0953}.

\bibitem{Menanteau2013}
F.~{Menanteau}, C.~{Sif{\'o}n}, L.~F. {Barrientos} et~al.
\newblock \emph{The Atacama Cosmology Telescope: Physical Properties of
  Sunyaev-Zel'dovich Effect Clusters on the Celestial Equator}.
\newblock \emph{\apj} \textbf{765} (2013) 67.
\newblock \eprint{1210.4048}.

\bibitem{Hotchkiss2011}
S.~{Hotchkiss}.
\newblock \emph{Quantifying the rareness of extreme galaxy clusters}.
\newblock \emph{\jcap} \textbf{7} (2011) 4.
\newblock \eprint{1105.3630}.

\bibitem{Hoyle2012}
B.~{Hoyle}, R.~{Jimenez}, L.~{Verde} et~al.
\newblock \emph{A critical analysis of high-redshift, massive, galaxy clusters.
  Part I}.
\newblock \emph{\jcap} \textbf{2} (2012) 9.
\newblock \eprint{1108.5458}.

\bibitem{Komatsu2011}
E.~{Komatsu}, K.~M. {Smith}, J.~{Dunkley} et~al.
\newblock \emph{Seven-year Wilkinson Microwave Anisotropy Probe (WMAP)
  Observations: Cosmological Interpretation}.
\newblock \emph{\apjs} \textbf{192} (2011) 18.
\newblock \eprint{1001.4538}.

\bibitem{Tinker2008}
J.~{Tinker}, A.~V. {Kravtsov}, A.~{Klypin} et~al.
\newblock \emph{Toward a Halo Mass Function for Precision Cosmology: The Limits
  of Universality}.
\newblock \emph{\apj} \textbf{688} (2008) 709.
\newblock \eprint{0803.2706}.

\bibitem{Mortonson2011}
M.~J. {Mortonson}, W.~{Hu} and D.~{Huterer}.
\newblock \emph{Simultaneous falsification of {$\Lambda$}CDM and quintessence
  with massive, distant clusters}.
\newblock \emph{\prd} \textbf{83}~(2) (2011) 023015.
\newblock \eprint{1011.0004}.

\bibitem{Andreon2009}
S.~{Andreon}.
\newblock \emph{Bayesian Methods in Cosmology}, Cambridge, chapter~12, p. 268
  (2009).

\bibitem{Stalder2013}
B.~{Stalder}, J.~{Ruel}, R.~{{\v S}uhada} et~al.
\newblock \emph{SPT-CL J0205-5829: A z = 1.32 Evolved Massive Galaxy Cluster in
  the South Pole Telescope Sunyaev-Zel'dovich Effect Survey}.
\newblock \emph{\apj} \textbf{763} (2013) 93.
\newblock \eprint{1205.6478}.

\bibitem{Rozo2012}
E.~{Rozo}, J.~G. {Bartlett}, A.~E. {Evrard} et~al.
\newblock \emph{Closing the Loop: A Self-Consistent Model of Optical, X-ray,
  and SZ Scaling Relations for Clusters of Galaxies}.
\newblock \emph{ArXiv e-prints} \eprint{1204.6305}.

\bibitem{Angulo2012}
R.~E. {Angulo}, V.~{Springel}, S.~D.~M. {White} et~al.
\newblock \emph{Scaling relations for galaxy clusters in the Millennium-XXL
  simulation}.
\newblock \emph{\mnras} \textbf{426} (2012) 2046.
\newblock \eprint{1203.3216}.

\bibitem{Piffaretti2011}
R.~{Piffaretti}, M.~{Arnaud}, G.~W. {Pratt} et~al.
\newblock \emph{The MCXC: a meta-catalogue of x-ray detected clusters of
  galaxies}.
\newblock \emph{\aap} \textbf{534} (2011) A109.
\newblock \eprint{1007.1916}.

\bibitem{Duffy2008}
A.~R. {Duffy}, J.~{Schaye}, S.~T. {Kay} et~al.
\newblock \emph{Dark matter halo concentrations in the Wilkinson Microwave
  Anisotropy Probe year 5 cosmology}.
\newblock \emph{\mnras} \textbf{390} (2008) L64.
\newblock \eprint{0804.2486}.

\bibitem{McInnes2009}
R.~N. {McInnes}, F.~{Menanteau}, A.~F. {Heavens} et~al.
\newblock \emph{First lensing measurements of SZ-detected clusters}.
\newblock \emph{\mnras} \textbf{399} (2009) L84.
\newblock \eprint{0903.4410}.

\bibitem{High2010}
F.~W. {High}, B.~{Stalder}, J.~{Song} et~al.
\newblock \emph{Optical Redshift and Richness Estimates for Galaxy Clusters
  Selected with the Sunyaev-Zel'dovich Effect from 2008 South Pole Telescope
  Observations}.
\newblock \emph{\apj} \textbf{723} (2010) 1736.
\newblock \eprint{1003.0005}.

\bibitem{vSuhada2010}
R.~{{\v S}uhada}, J.~{Song}, H.~{B{\"o}hringer} et~al.
\newblock \emph{XMM-Newton detection of two clusters of galaxies with strong
  SPT Sunyaev-Zel'dovich effect signatures}.
\newblock \emph{\aap} \textbf{514} (2010) L3.
\newblock \eprint{1003.3020}.

\bibitem{Andersson2011}
K.~{Andersson}, B.~A. {Benson}, P.~A.~R. {Ade} et~al.
\newblock \emph{X-Ray Properties of the First Sunyaev-Zel'dovich Effect
  Selected Galaxy Cluster Sample from the South Pole Telescope}.
\newblock \emph{\apj} \textbf{738} (2011) 48.
\newblock \eprint{1006.3068}.

\bibitem{Brodwin2010}
M.~{Brodwin}, J.~{Ruel}, P.~A.~R. {Ade} et~al.
\newblock \emph{SPT-CL J0546-5345: A Massive $z{>}1$ Galaxy Cluster Selected
  Via the Sunyaev-Zel'dovich Effect with the South Pole Telescope}.
\newblock \emph{\apj} \textbf{721} (2010) 90.
\newblock \eprint{1006.5639}.

\bibitem{Marriage2011}
T.~A. {Marriage}, V.~{Acquaviva}, P.~A.~R. {Ade} et~al.
\newblock \emph{The Atacama Cosmology Telescope: Sunyaev-Zel'dovich-Selected
  Galaxy Clusters at 148 GHz in the 2008 Survey}.
\newblock \emph{\apj} \textbf{737} (2011) 61.
\newblock \eprint{1010.1065}.

\bibitem{Foley2011}
R.~J. {Foley}, K.~{Andersson}, G.~{Bazin} et~al.
\newblock \emph{Discovery and Cosmological Implications of SPT-CL J2106-5844,
  the Most Massive Known Cluster at $z{>}1$}.
\newblock \emph{\apj} \textbf{731} (2011) 86.
\newblock \eprint{1101.1286}.

\bibitem{ThePlanckCollaboration2011}
{The Planck Collaboration}, N.~{Aghanim}, M.~{Arnaud} et~al.
\newblock \emph{Planck Intermediate Results. I. Further validation of new
  Planck clusters with XMM-Newton}.
\newblock \emph{ArXiv e-prints} \eprint{1112.5595}.

\bibitem{Reichardt2013}
C.~L. {Reichardt}, B.~{Stalder}, L.~E. {Bleem} et~al.
\newblock \emph{Galaxy Clusters Discovered via the Sunyaev-Zel'dovich Effect in
  the First 720 Square Degrees of the South Pole Telescope Survey}.
\newblock \emph{\apj} \textbf{763} (2013) 127.
\newblock \eprint{1203.5775}.

\bibitem{High2012}
F.~W. {High}, H.~{Hoekstra}, N.~{Leethochawalit} et~al.
\newblock \emph{Weak-lensing Mass Measurements of Five Galaxy Clusters in the
  South Pole Telescope Survey Using Magellan/Megacam}.
\newblock \emph{\apj} \textbf{758} (2012) 68.
\newblock \eprint{1205.3103}.

\bibitem{Consortium2012}
A.~{Consortium}, {:}, M.~P. {Schammel} et~al.
\newblock \emph{AMI SZ observations and Bayesian analysis of a sample of six
  redshift-one clusters of galaxies}.
\newblock \emph{ArXiv e-prints} \eprint{1210.7771}.

\bibitem{Hasselfield2013}
M.~{Hasselfield}, M.~{Hilton}, T.~A. {Marriage} et~al.
\newblock \emph{The Atacama Cosmology Telescope: Sunyaev-Zel'dovich Selected
  Galaxy Clusters at 148 GHz from Three Seasons of Data}.
\newblock \emph{ArXiv e-prints} \eprint{1301.0816}.

\bibitem{Rosati2009}
P.~{Rosati}, P.~{Tozzi}, R.~{Gobat} et~al.
\newblock \emph{Multi-wavelength study of XMMU J2235.3-2557: the most massive
  galaxy cluster at $z {>} 1$}.
\newblock \emph{\aap} \textbf{508} (2009) 583.
\newblock \eprint{0910.1716}.

\bibitem{Gobat2011}
R.~{Gobat}, E.~{Daddi}, M.~{Onodera} et~al.
\newblock \emph{A mature cluster with X-ray emission at z = 2.07}.
\newblock \emph{\aap} \textbf{526} (2011) A133.
\newblock \eprint{1011.1837}.

\bibitem{Fassbender2011}
R.~{Fassbender}, H.~{B{\"o}hringer}, A.~{Nastasi} et~al.
\newblock \emph{The x-ray luminous galaxy cluster population at $0.9 {<} z
  {\lesssim} 1.6$ as revealed by the XMM-Newton Distant Cluster Project}.
\newblock \emph{New Journal of Physics} \textbf{13}~(12) (2011) 125014.
\newblock \eprint{1111.0009}.

\bibitem{vSuhada2012}
R.~{{\v S}uhada}, J.~{Song}, H.~{B{\"o}hringer} et~al.
\newblock \emph{The XMM-BCS galaxy cluster survey. I. The X-ray selected
  cluster catalog from the initial 6 deg$^{2}$}.
\newblock \emph{\aap} \textbf{537} (2012) A39.
\newblock \eprint{1111.0141}.

\bibitem{Okabe2010}
N.~{Okabe}, M.~{Takada}, K.~{Umetsu} et~al.
\newblock \emph{LoCuSS: Subaru Weak Lensing Study of 30 Galaxy Clusters}.
\newblock \emph{\pasj} \textbf{62} (2010) 811.
\newblock \eprint{0903.1103}.

\bibitem{Demarco2010}
R.~{Demarco}, G.~{Wilson}, A.~{Muzzin} et~al.
\newblock \emph{Spectroscopic Confirmation of Three Red-sequence Selected
  Galaxy Clusters at z = 0.87, 1.16, and 1.21 from the SpARCS Survey}.
\newblock \emph{\apj} \textbf{711} (2010) 1185.
\newblock \eprint{1002.0160}.

\bibitem{Menanteau2010}
F.~{Menanteau}, J.~P. {Hughes}, L.~F. {Barrientos} et~al.
\newblock \emph{Southern Cosmology Survey. II. Massive Optically Selected
  Clusters from 70 Square Degrees of the Sunyaev-Zel'dovich Effect Common
  Survey Area}.
\newblock \emph{\apjs} \textbf{191} (2010) 340.
\newblock \eprint{1002.2226}.

\bibitem{Vulcani2012}
B.~{Vulcani}, A.~{Arag{\'o}n-Salamanca}, B.~M. {Poggianti} et~al.
\newblock \emph{Cl 1103.7-1245 at z = 0.96: the highest redshift galaxy cluster
  in the EDisCS survey}.
\newblock \emph{\aap} \textbf{544} (2012) A104.
\newblock \eprint{1207.1530}.

\bibitem{Davis2011}
O.~{Davis}, J.~{Devriendt}, S.~{Colombi} et~al.
\newblock \emph{Most massive haloes with Gumbel statistics}.
\newblock \emph{\mnras} \textbf{413} (2011) 2087.
\newblock \eprint{1101.2896}.

\bibitem{Gumbel1958}
E.~J. Gumbel.
\newblock \emph{Statistics of Extremes}.
\newblock Columbia University Press (1958).

\end{thebibliography}
\end{document}